\documentclass[12pt,a4paper]{article}
\usepackage{amsmath,amssymb,amsfonts,epsfig,graphicx}

\newcommand{\be}{\begin{equation}}
\newcommand{\bea}{\begin{eqnarray}}
\newcommand{\eea}{\end{eqnarray}}
\newcommand{\ba}{\begin{array}}
\newcommand{\ea}{\end{array}}
\newcommand{\ee}{\end{equation}}
\begin{document}
\begin{titlepage}
\hfill \vbox{
    \halign{#\hfil         \cr
           IPM/P-2007/033   \cr
           arXiv:0705.3131 [hep-th] \cr
           } 
      }  
\vspace*{6mm}
\begin{center}
{\Large {\bf Dual Spikes}}\\
\vspace*{3mm}
{\large{\bf New Spiky String Solutions}} \vspace*{10mm}
\vspace*{1mm}

 {\bf  A. E. Mosaffa$^{1}$, B. Safarzadeh$^{{1},{2}}$}

\vspace*{0.4cm}

{\it ${}^1${Institute for Studies in Theoretical Physics and Mathematics (IPM)\\
P.O.Box 19395-5531, Tehran, Iran}}

\vspace*{0.4cm}

{\it ${}^2${Department of physics, School of Science Tarbiat Modares
University, P.O. Box 14155-4838, Tehran, Iran }}

\vspace*{0.2cm}

{\tt mosaffa, safarzadeh@theory.ipm.ac.ir }

\vspace*{2cm}
\end{center}

\begin{center}
{\bf\large Abstract}
\end{center}

We find a new class of spiky solutions for closed strings in flat, $AdS_3\subset AdS_5$
and $R\times S^2(\subset S^5)$
backgrounds. In the flat case the new solutions turn out to be T-dual configurations of spiky strings
found in \cite{Kkk}. In the case of solutions living in $AdS$, we make a semi classical analysis by
taking the large angular momentum limit. The anomalous dimension for these dual spikes is similar to that
for rotating and pulsating circular strings in AdS with angular momentum playing the role of the level number.
This replaces the well known logarithmic dependence for spinning
strings. For the dual spikes living on sphere we find that no large
angular momentum limit exists.

\end{titlepage}

\section{Introduction}
The AdS/CFT dictionary is far from complete, the main obstacle being the strong/weak nature of
Maldacena's conjecture\cite{Malda},\cite{Gubser},\cite{Witten}.
In an important line of research in this context, which has been widely studied,
certain regimes and limits have been considered where one can study both sides
of the duality perturbatively. The main idea for this has been focusing on
states/operators with large charges \cite{BMN},\cite{GKP}.

A large charge, $J$,  introduces a free parameter, besides the 't Hooft coupling $\lambda$,
into the game that can give way for some control
over the problem. One can then define an effective 't Hooft coupling, $\lambda'$,
built from $\lambda$ and $J$, which is held fixed in the large charge limit.
The idea is that the world sheet loop expansions around certain classical
string solutions with a properly chosen $J$, are generically suppressed by inverse powers of $J$
\footnote{For this to happen one actually needs at least one large charge on $S^5$. For charges
only inside AdS, the semi classical expressions are only reliable when $\lambda\gg1$ but it is expected,
verified by some tests, that for large charges these expressions
can be extrapolated to perturbative field theory results
for anomalous dimensions of the conjectured corresponding operators.}.
In SYM this translates into the distinction of a class of SYM operators,
determined by the charges, whose couplings to the rest of the operators are
suppressed by inverse powers of $J$.

The best known example in this context is the BMN case \cite{BMN}
(see e.g. \cite{Sadri} for reviews). In the strict BMN limit,
the world sheet expansion around
a certain point-like string solution in $AdS_5\times S^5$ terminates
at one loop and the corresponding class of SYM operators consists only of the BMN operators,
decoupled from the rest. Moreover, the one loop action can be solved exactly \cite{7ppwave}
with a closed expression for the spectrum of fluctuations around the
classical configuration. This can be matched order by order with a perturbative
calculation for the anomalous dimension of BMN operators which can be done when
 $\lambda'\equiv \lambda/J^2\ll1$.

In some other interesting cases, adding more large charges
can even make the one loop world sheet expansion
sub leading and one is left with only the classical expression for the energy of
the string configuration. That is, the classical expression gives an all loop quantum
prediction for the field theory operators in a certain class.
This is the case for string solutions with spin and angular
momentum.

In another significant development, perturbative planar calculations for
anomalous dimensions of SYM operators led to the discovery of an integrable structure
in the system \cite{Minahan:2002ve}. This came through the identification
of the dilatation operator (see \cite{Beisert:2002ff}), whose eigenvalues are the scaling dimensions of
SYM operators, with the Hamiltonian of an equivalent spin chain. The problem then
boils down to diagonalizing the Hamiltonian by solving
a set of Bethe equations for the spin chain. This method has had some successes even beyond
semi classical limit and large charges, which define infinitely long chains.
The integrability
structure in the string theory side has also been discovered \cite{Mandal:2002fs} and further studied
in the semi classical limit \cite{Arutyunov:2003uj}. For semi classical analysis of strings
and spin chain developments see e.g. \cite{Frolov:2002av}, \cite{spinchain}.
For reviews and references on these subjects see e.g. \cite{semireviews}.

In \cite{Kkk}
a new class of string solutions in AdS space
was found with a number of spikes on the string\footnote{Spiky strings in flat space
as cosmic strings were studied in \cite{Burden:1985md}.}. These
spiky strings were identified with higher twist operators in SYM with
each spike representing a particle in the field theory. Large angular
momentum is provided by a large number of covariant derivatives acting on the fields which
produce the mentioned particles. The total number of derivatives is
distributed equally amongst the fields for these solutions. Spiky strings
were further generalized to configurations on sphere \cite{Ryang}. Certain limits of these
solutions were shown \cite{KRT} to correspond to giant magnons \cite{HM} which
represent spin waves with a short wave length in the spin chain language. Solutions
corresponding to multi spin giant magnons  \cite{Bobev:2006fg} and those
with a magnon like dispersion relation in M-theory \cite{Bozhilov:2006bi} were also 
studied.

In this paper we generalize the spiky string solutions in both AdS and sphere spaces.
These solutions, which we will call ``dual spikes", represent spiky strings with
the direction of spikes reversed. In flat space, these dual spikes are shown to be T-dual to
the usual ones. In AdS we will study the large angular momentum limit of the solution
and derive its energy, $E$, in terms of angular momentum, $J$. We find that these dual spikes
represent higher twist operators whose anomalous dimensions are similar to those of
rotating and pulsating circular strings. That is, the anomalous dimension in
the large $J$ limit is proportional to $\lambda^{1/4}\sqrt{J}$. This replaces
the usual logarithm dependence for folded and spiky spinning strings in AdS.
We will argue that this is an expected behavior, as for fast spinning dual spikes
and near the boundary, we are effectively dealing with portions of almost circular
strings with a pulsation-like motion which is induced by a profile of string
in the AdS radius and angular momentum. One might then conclude that the
corresponding operators for such configurations, in addition to a large
number of covariant derivatives, $D_+$, which induce spin in $S^3\subset AdS_5$,
also contain the combination $D_+ D_-$ which induce an effective pulsation.
This last combination which contributes to the scaling dimension but not
to spin could be responsible for a change from logarithm dependence
to square root in the anomalous part.

In $S^5$ however, we will
show that unlike the usual spikes, the dual spikes have no large angular momentum limit.
We will find the $E-J$ of nearly circular strings for the cases that the
string lives near the pole or near the equator.\\

{\large\bf{Note Added:\ }}
 While this work was being prepared we learned of a related work on dual spikes which
appeared very recently \cite{Ishizeki:2007we}.


\section{Spiky strings and their T-duals in flat space-time}

In this section we find a two parameter family of solutions describing closed strings with spikes in flat space. These
solutions fall into two distinct classes depending on whether the ratio of the two parameters is greater or
smaller than one. One of the two classes describes the spiky string solutions found in \cite{Kkk} and the the other one,
as we will see in what follows, can be obtained from the first one by a  T-duality transformation.

Let us start with the Nambu-Gotto action and the following
ansatz for the string

\begin{equation}
\label{ansatz}
t=\tau, \hspace{1cm} \theta=\omega\ \tau+\sigma,
\hspace{1cm}r=r(\sigma).
\end{equation}

where $(\tau,\sigma)$ and $(t,r,\theta)$ are the world sheet and target space coordinates respectively
and $r_c$ is a constant. For the flat target space

\begin{equation}
ds^2=-dt^2+r^2+r^2d\theta^2
\end{equation}

and the ansatz (\ref{ansatz}), the NG action is found as

\begin{equation}
{\cal L}_{NG}=-\sqrt{l}\ ,\ \ \ \ \ l=(1-\omega^2\ r^2)\ r'^2+r^2
\end{equation}

where prime denotes $\partial_{\sigma}$. The constant of motion, $r_l$, associated with the
$\partial_{\theta}$ isometry is

\[
\frac{r^2}{\sqrt{l}}\equiv r_l
\]

Plugging this constant in the equations of motion gives

\be
\label{rel}
\frac{r'^2}{r^2}=\frac{r_c^2}{r_l^2}\ \frac{r^2-r_l^2}{r_c^2-r^2}
\ee

where $r_c=1/\omega$. One can show that $r(\sigma)$ found from (\ref{rel}) satisfies the equations of motion.
The two constants $r_l$ and $r_c$ parameterize the solutions and, as can be seen from (\ref{rel}), they correspond to the
radius of the lobe and cusp of the string respectively. Defining

\[
r_1=min(r_c,r_l)\ , \ \ \ \ \ \ \ \ \ \ \ \  r_2=max(r_c,r_l)
\]

we can write the expressions for the energy, $E$, and angular momentum, $J$, of a segment of string
stretched between $r_1$ and $r_2$

\begin{eqnarray}
E_{seg}&=&\frac{1}{2\pi}\frac{1}{r_c}\int_{r_1}^{r_2} dr r \frac{|r_c^2-r_l^2|}{\sqrt{(r^2-r_l^2)(r_c^2-r^2)}}
=\frac{1}{4}\frac{r_c}{a^2}|a^2-1|\nonumber\\
J_{seg}&=&=\frac{1}{2\pi}\int_{r_1}^{r_2} dr r \sqrt{\frac{r^2-r_l^2}{r_c^2-r^2}}=\frac{1}{8}\frac{r_c^2}{a^2}|a^2-1|
\end{eqnarray}

where we have defined the constant $a$ by

\[
a\equiv\frac{r_c}{r_l}
\]

The angle covered by the segment, $\Delta\theta$, can also be found as

\be
\Delta\theta=\frac{r_l}{r_c}\int_{r_1}^{r_2}\frac{dr}{r}\sqrt{\frac{r_c^2-r^2}{r_2-r_l^2}}=\frac{\pi}{2}\frac{|a-1|}{a}
\ee

We now demand that $2n$ number of segments make up a closed string and hence we have $n$ number of spikes
on the string. This will give

\be
\label{cons}
\frac{2a}{|a-1|}=n
\ee

As a result, the total energy and angular momentum of the string will be found as

\be
E=\frac{r_c}{a}(a+1)\ , \ \ \ \ \ \ \ \ \ \ \ \ J=\frac{r_c^2}{2a}(a+1)
\ee
\newpage
with the obvious relation

\be
\label{EJ}
E=2\frac{J}{r_c}
\ee

Therefore to each pair, $(a,r_c)$, corresponds a unique string configuration provided that the periodicity condition
(\ref{cons}) is satisfied for some integer $n$.
We can now identify two distinct cases; $a>1$ and $a<1$. It is readily seen that the first case
describes the spiky solutions found in \cite{Kkk} with the following
relations

\be
a=\frac{n}{n-2}\ ,\ \ \ \ \ \ \ \ \ \ E=2r_c\frac{n-1}{n}\ ,\ \ \ \ \ \ \ \ \ \ J=r_c^2\ \frac{n-1}{n}
\ee

and the dispersion relation

\be
E=2\sqrt{\frac{n-1}{n}J}
\ee

For the second case we have the following new relations

\be
a=\frac{n}{n+2}\ ,\ \ \ \ \ \ \ \ \ \ E=2r_c\frac{n+1}{n}\ ,\ \ \ \ \ \ \ \ \ \ J=r_c^2\frac{n+1}{n}
\ee

and the charges are related as

\be
E=2\sqrt{\frac{n+1}{n}J}
\ee

One can easily check that the energy determined by the pair $(a,r_c)$ remains invariant if we switch
to the pair $(1/a,r_c/a)$. Equivalently this amounts to interchanging $r_c$ and $r_l$.
Under such a transformation $J$ goes over to $J/a$ such
that the dispersion relation (\ref{EJ}) also remains invariant.
As can be seen from (\ref{cons}), this transformation takes an $n$ spike configuration in the first class
to an $n-2$ spike configuration in the second with the same energy and vice versa.
The whole set of transformations is thus as follows

\bea
\label{Tdual}
a&\rightarrow& \frac{1}{a}\nonumber\\
r_c&\rightarrow&\frac{r_c}{a}\nonumber\\
n&\rightarrow&n+2\frac{|1-a|}{1-a}\\
J&\rightarrow&\frac{J}{a}\nonumber\\
E&\rightarrow&E\nonumber
\eea

\begin{figure}[ht]
\ \ \ \ \ \ \ \ \includegraphics[scale=0.7]{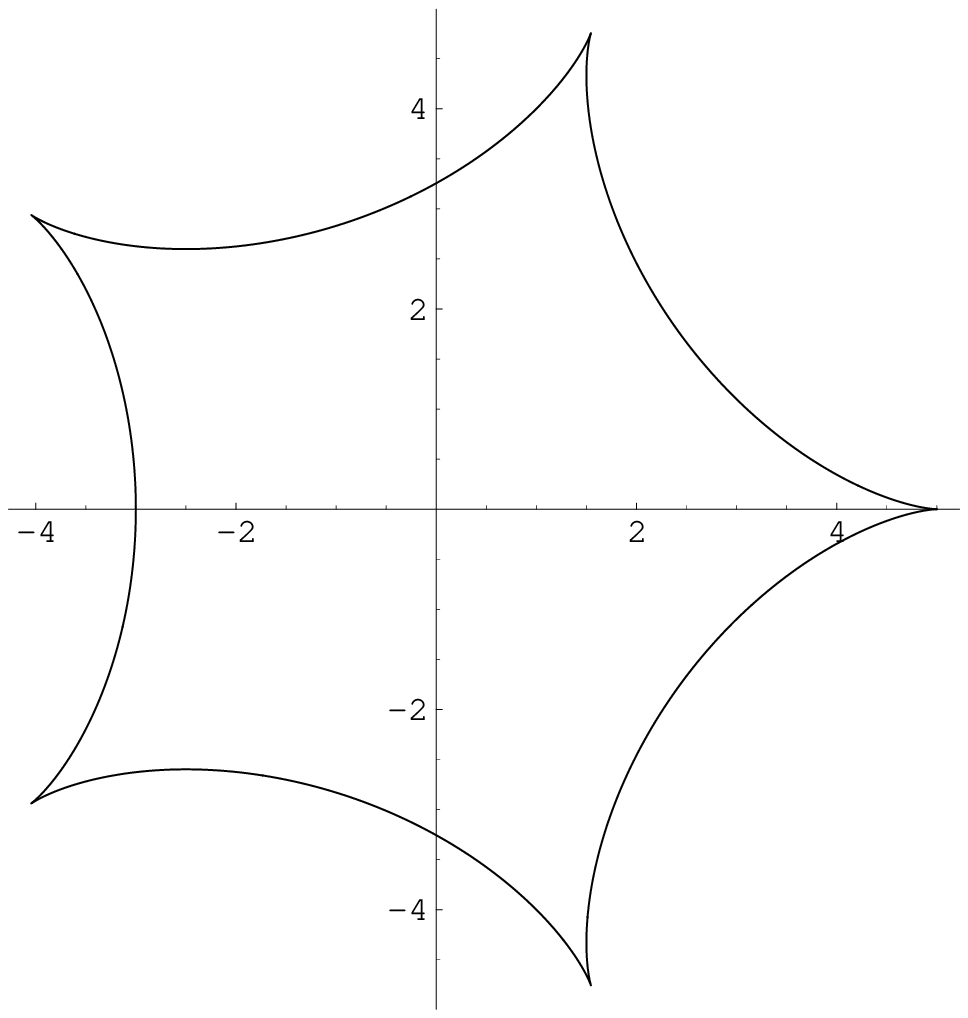}\ \ \ \ \ \ \ \ \ \ \ \ \ \ \ \ \
\includegraphics[scale=0.7]{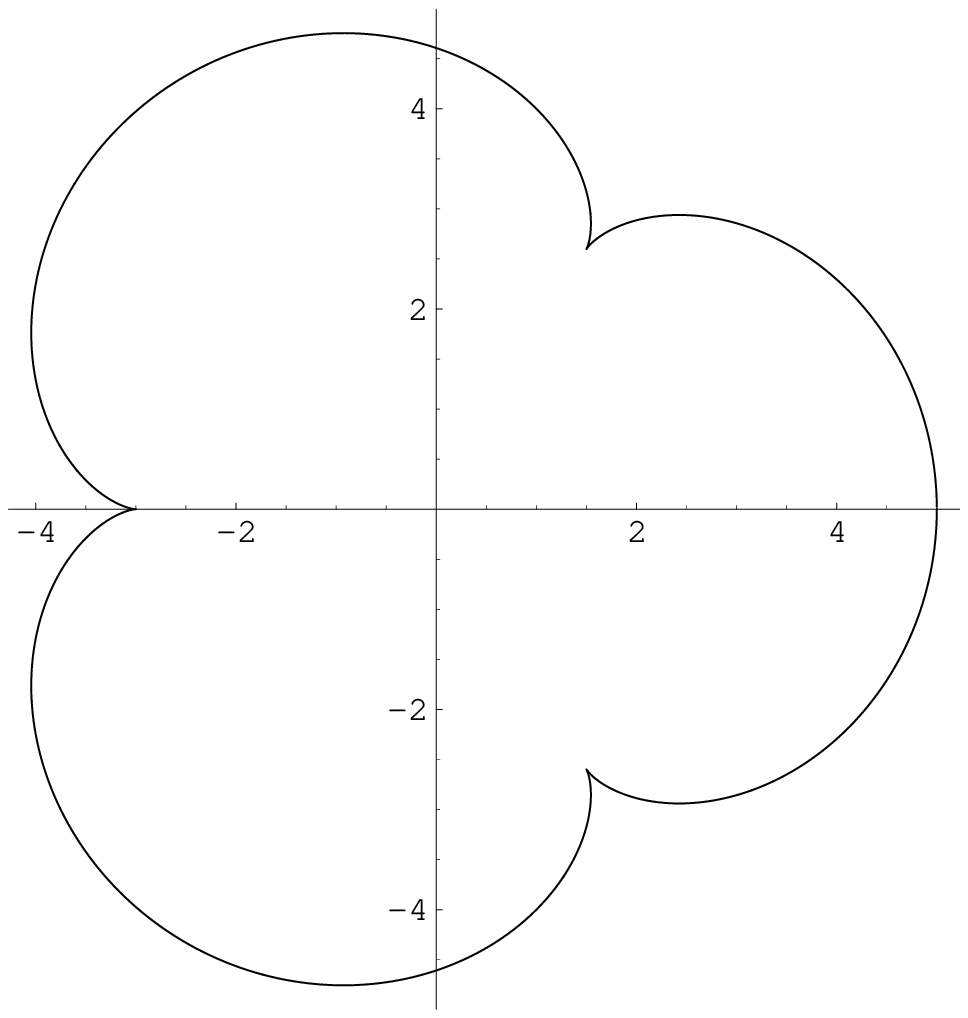}
\caption{Spiky strings related by T duality. The pair $(a,r_c)$ for the figure on the left
is $(5/3,5)$ which gives a $5$ spike configuration whereas for the one on the right is $(3/5,3)$ and results
in a $3$ spike solution}
\label{5to3}
\end{figure}

For a finite $a>1$, we have a closed string with $n$ spikes pointing outwards, $n$ determined by (\ref{cons}).
As $a$ decreases,
the number of spikes on the string increases until it approaches a circle which is the limiting
string configuration for this class of solutions. This happens when $a=1$ and $n=\infty$.
The complementary range of values for $a$, $0<a<1$, produces
configurations for the string which can be obtained from the first class by the transformations in (\ref{Tdual}).
This time, however, the spikes point inwards and as we decrease $a$, the number of spikes on the string also decreases from $n=\infty$
(a circular string) to $n=1$.

The limiting value of $a\rightarrow\infty$ corresponds to
a folded string with $n=2$ which rotates in a circle of radius $r_c$ with angular momentum $\omega=1/r_c$.
This configuration does not have a corresponding dual in the second class as the string segment
for $a=0$ describes a spiral stretching from $r_c=0$ to $r_l$ which covers an infinite $\Delta\theta$ and
has an infinite energy\footnote{In a recent paper \cite{Ishizeki:2007we}, such configurations
have been studied on sphere with the name ``single spikes".}. So we demand that in the second family $a$ is at least
equal to $1/3$ which is the minimum value that can produce a closed string with only one spike.
In short, one can have spiky configurations with $n\ge2$ for $a>1$ and
$n\ge1$ for $1/3\le a<1$. The $a=1$ case describes a circle and is self dual.

\begin{figure}[ht]
\ \ \ \ \ \ \ \ \ \includegraphics[scale=0.7]{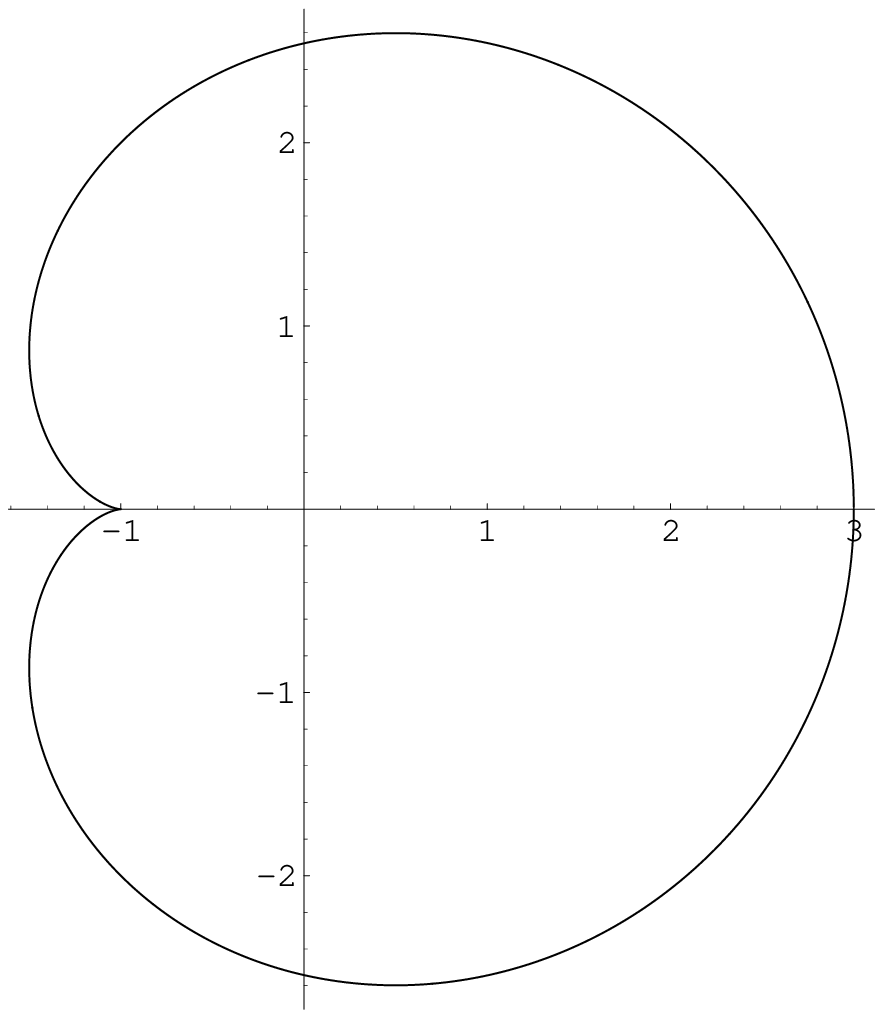}\ \ \ \ \ \ \ \ \ \ \ \ \ \ \ \
\includegraphics[scale=0.7]{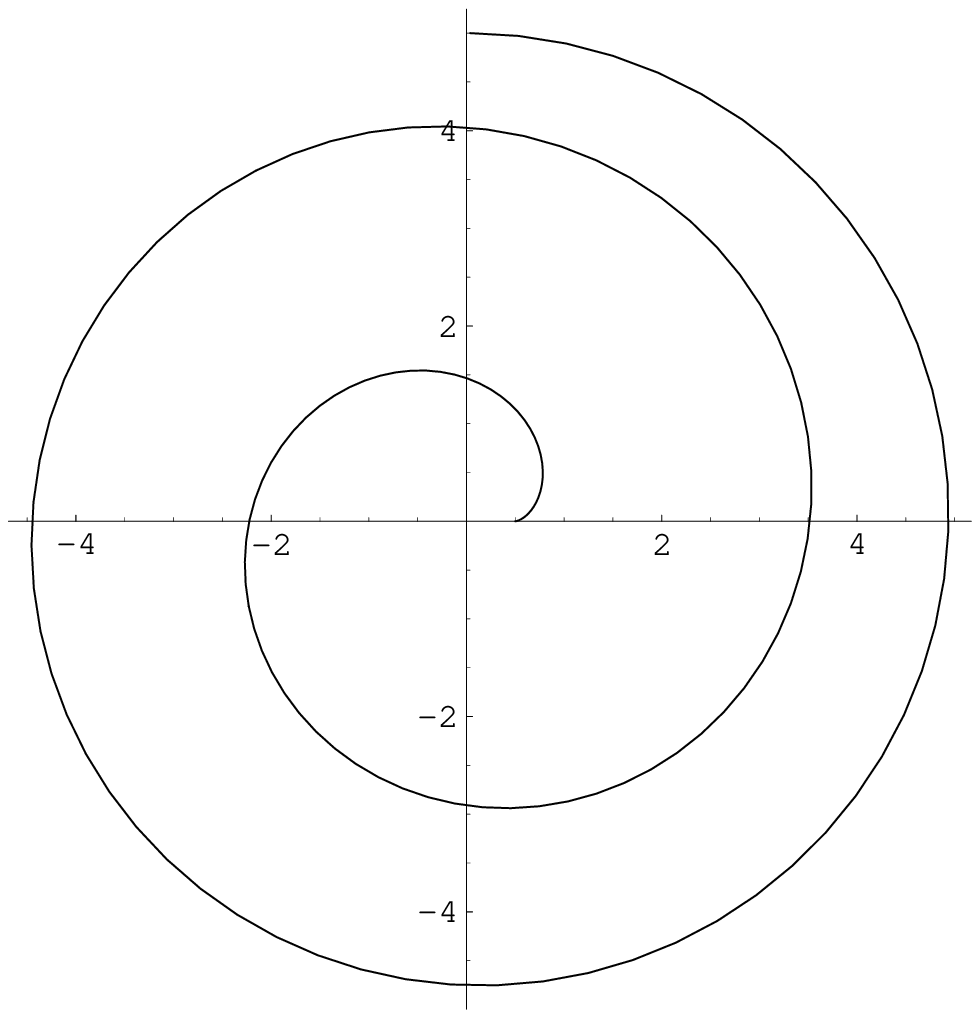}
\caption{The figure on the left is a dual closed spiky string with minimum number of spikes, $n=1$,
which is given by $a=1/3$. The figure on the right is a segment of string stretching between $r_c$ and $r_l$
with $a<1(a=0.1$ here).}
\end{figure}

One can show that (\ref{Tdual}) is in fact a T-duality
transformation in the $(r,\theta)$ plane.
To see this, we find the above spiky solutions using the Polyakov action for the string.
The following configuration is by construction a solution to the wave equations resulting from this action

\begin{eqnarray}
\label{flat}
\label{pol}
x&=&r_c\frac{|a-1|}{2a} \cos(\frac{a+1}{a-1}\sigma_+)+r_c\frac{a+1}{2a}\cos(\sigma_{-})\nonumber\\
y&=&r_c\frac{|a-1|}{2a} \sin(\frac{a+1}{a-1}\sigma_+)+r_c\frac{a+1}{2a}\sin(\sigma_{-})\\
t&=&r_c\frac{a+1}{a}\tau\nonumber
\end{eqnarray}

where $\sigma_+= \tau+\sigma$, $\sigma_-= \tau-\sigma$,
and the target space coordinates $(t,x,y)$ parameterize
a Minkowski space. Note that there are two free parameters in this solution, $a$ and $r_c$.
We further demand that the relation (\ref{cons}) holds for some $n$.
This guarantees, as can be easily checked, that (\ref{flat}) satisfies periodic boundary condition
as well as the level matching and Virasoro constraints. Therefore we find a two
parameter family of closed string solutions determined by $(a,r_c)$.

Each pair $(a,r_c)$, together with (\ref{cons}), describes a spiky closed string with
$n$ spikes in conformal gauge.
The condition $X'^2=0$, where $X=(t,x,y)$, determines the position of cusps. This
gives the radius of spikes at $r_c$. The condition $X.X'=0$, on the other hand,
gives the position of lobes which results in $r_c/a\equiv r_l$ for the radius of lobes.

For $a>1$, the solutions (\ref{flat}) are those found in \cite{Kkk}. For $1/3\le a<1$
we recover the second class of spiky solutions discussed above. Applying the
transformations (\ref{Tdual}) to a given solution, i.e. $(a,r_c)\rightarrow(1/a,r_c/a)$, amounts to
changing the sign of the left mover part of $y$, $y_L$. This transformation, $y_L\rightarrow -y_L$,
is of course T-duality in $y$ direction. We hence conclude that, as mentioned above, (\ref{Tdual})
is a T-duality transformation on solutions.


\section{Dual spikes in $AdS$}

In this section we find spiky solutions in the $AdS$ background whose spikes
point inward, like the T-dual spikes found in the previous section. This is a
generalization of spiky strings in $AdS$ with outward spikes found in \cite{Kkk}.

Our main interest is the $AdS_5\times S_5$ background, in the global coordinates, but the closed string we are
interested in lives in the $AdS_3$ subspace specified by the following metric

\be
ds^2=-\cosh^2\rho dt^2+d\rho^2+\sinh^2\rho d\theta^2
\ee

\newpage

Our ansatz for the string configuration is the following

\be
t=\tau\ ,\ \ \ \ \ \theta=\omega\tau+\sigma\ ,\ \ \ \ \ \ \rho=\rho(\sigma)
\ee

The radius of $AdS$ is chosen to be one and the dimensionless worldsheet
coupling constant is denoted by $1/\sqrt{\lambda}$ where it is understood that
$\lambda$ is the 't Hooft coupling in the dual field theory, $N=4$ SYM. The
Nambu-Gotto action for the above ansatz is

\[
{\cal L}_{NG}=-\frac{\sqrt{\lambda}}{2\pi}\sqrt{l}\nonumber
\]

where

\be
l=(\cosh^2\rho-\omega^2\sinh^2\rho)\rho'^2+\sinh^2\rho\cosh^2\rho
\ee

The isometry direction, $\partial_{\theta}$, results in a constant of motion, which as
we see below, determines the position of lobes in the string configuration and hence we denote
it by ${\rho_l}$

\be
\label{lobe}
\frac{\sinh2\rho_l}{2}\equiv\frac{\sinh^2\rho\ \cosh^2\rho}{\sqrt{l}}
\ee

The other free parameter of the problem, $\omega$, on the other hand, determines
the position of cusps, $\rho_c$, which we define by

\be
\label{cusp}
\sinh^2{\rho_c}\equiv\frac{1}{\omega^2-1}
\ee

From (\ref{lobe}) it follows that

\be
\label{integral}
\frac{\rho'}{\sinh2\rho}=\frac{1}{2}\frac{\sqrt{\cosh2\rho_c-1}}{\sinh2\rho_l}
\sqrt{\frac{\cosh^22\rho-\cosh^22\rho_l}{\cosh2\rho_c-\cosh2\rho}}
\ee

One can check that if the above relation holds, the equations of motion are also satisfied.
It is readily seen that $\rho_l$ and $\rho_c$ indeed determine the position of lobes and cusps
on the string respectively.

All the above relations were found in \cite{Kkk}. What we do in the following is to change the assumption
made in \cite{Kkk}, namely $\rho_c>\rho_l$, to $\rho_c<\rho_l$ and find a whole new family
of solutions which, unlike their counterparts in the mentioned reference, describe strings with
spikes pointing towards the origin of $AdS$.

To proceed, we make the useful change of variables mentioned in \cite{Kkk} from
$\rho$ to $u$ defined by $u=\cosh2\rho$. In terms of this new variable we can write
the following relations

\bea
\Delta\theta&=&\sqrt{\frac{u_l^2-1}{u_c-1}}\int_{u_c}^{u_l}\frac{du}{u^2-1}\frac{\sqrt{u-u_c}}{\sqrt{u_l^2-u^2}}\\
J&=&\frac{2n}{2\pi}\sqrt{\lambda}\frac{\sqrt{u_c-1}}{4}\int_{u_c}^{u_l}\frac{du}{u+1}\frac{\sqrt{u_l^2-u^2}}{\sqrt{u-u_c}}\\
E-\omega J&=&\frac{2n}{2\pi}\sqrt{\lambda}\frac{\sqrt{2}}{\sqrt{u_c-1}}\int_{u_c}^{u_l}du\frac{\sqrt{u-u_c}}{\sqrt{u_l^2-u^2}}
\eea

\begin{figure}[ht]
\begin{center}
\includegraphics[scale=0.85]{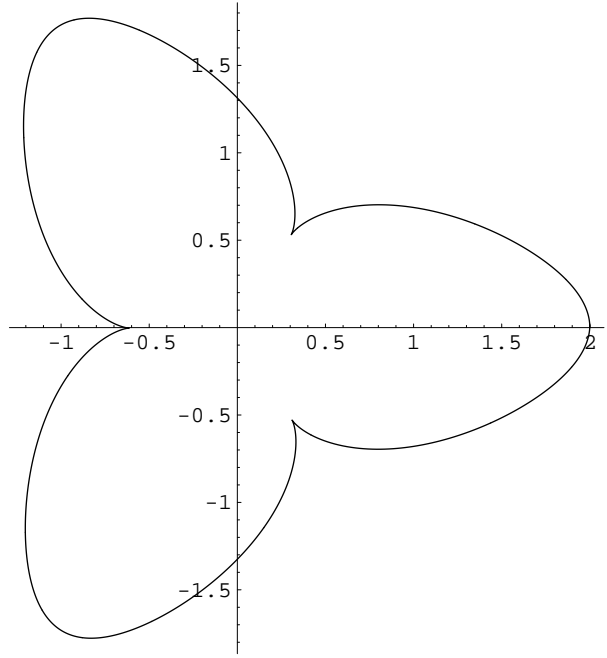}
\caption{A dual spiky string with $3$ spikes. $\rho_c=0.61347562$ and $\rho_l=2$.}
\end{center}
\end{figure}

In the above relations, each integration is performed on a segment of string stretching between
a cusp and a lobe. When we add up $2n$ number of such segments we make a closed string with $n$ cusps
provided that the angle covered by a segment satisfies $\Delta\theta=\pi/n$.

The integrals written above can be computed using integral tables (see e.g. \cite{grad}). The result is

\bea
\Delta\theta&=&\frac{\sqrt{u_l^2-1}}{\sqrt{2u_l(u_c-1)}}\ \{\frac{u_c+1}{u_l+1}\Pi(n_1,p)-\frac{u_c-1}{u_l-1}\Pi(n_2,p)\}\\
J&=&\frac{2n}{2\pi}\sqrt{\lambda}\frac{\sqrt{u_c+1}}{2\sqrt{2u_l}}\ \{(1+u_l)K(p)-2u_lE(p)+(u_l-1)\Pi(n_1,p)\}\\
E-\omega J&=&\frac{2n}{2\pi}\sqrt{\lambda}\frac{2\sqrt{u_l}}{\sqrt{u_c-1}}\ \{E(p)-\frac{u_c+u_l}{2u_l}K(p)\}
\eea

where

\be
n_1=\frac{u_l-u_c}{u_l+1}\ ,\ \ \ \ \ \ n_2=\frac{u_l-u_c}{u_l-1}\ ,\ \ \ \ \ \ p=\sqrt{\frac{u_l-u_c}{2u_l}}
\ee

and $K(p)$, $E(p)$ and $\Pi(n,p)$ are the complete elliptic integrals of first, second and third kind
respectively (the argument $n$ in $\Pi$ need not be an integer and shouldn't be confused with the number of spikes)(see Appendix A).

\begin{figure}[ht]
\begin{center}
\includegraphics[scale=0.85]{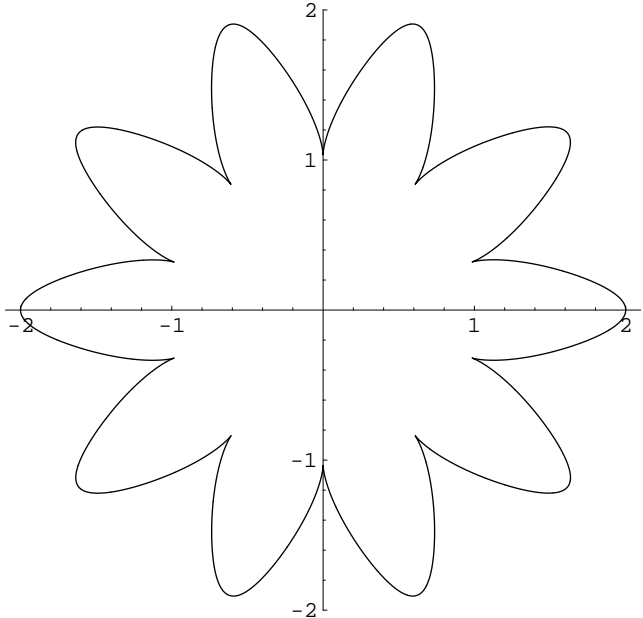}
\caption{A dual spiky string with $10$ spikes. $\rho_c=1.03593842$ and $\rho_l=2$.}
\end{center}
\end{figure}

The argument $p$ appearing in the elliptic integrals varies between zero and $1/\sqrt{2}$ and hence
$K(p)$ and $E(p)$ are always of order unity. The elliptic integrals of the third kind we are dealing with, $\Pi(n_i,p)$,
are ``circular" ones because of the relation $p^2<n_1,n_2<1$. These functions increase as $n_i$ and/or $p$ increase and
blow up as $n_i$ approaches $1$.

Due to the complicated form of the relations, and unlike the analysis made for the flat space,
we did not manage to make a direct and general comparison
between the two families of spiky solutions in $AdS$, i.e. the ones we have found here with inward spikes and
the ones found in \cite{Kkk}.
Therefore we study some limits where the solutions simplify and/or where
we may have a dual field theory description.

The first limit we consider is when $\rho_c<\rho_l\ll1$. This limit corresponds to
a small angular momentum spiky string close to the origin and reproduces the dual spikes found
in the previous section and all the relations there apply here.

The other limit, which for comparison with the dual field theory is of more interest, is when
the angular momentum is large. This happens when $\rho_l\gg1$ which we will study in the following in some detail.
So the limit we are interested in is when $\rho_c$ is fixed and $\rho_l\rightarrow\infty$.

The $\Pi(n_i,p)$ in this limit behave as

\[
\label{piapprox}
\Pi(n_1,p)\approx \frac{\pi}{2}\frac{\sqrt{2u_l}}{\sqrt{u_c-1}}\ ,\ \ \ \ \ \
\Pi(n_2,p)\approx \frac{\pi}{2}\frac{\sqrt{2u_l}}{\sqrt{u_c+1}}\nonumber
\]

In evaluating these limits we have made use of a relation which holds between the
circular elliptic integrals and the Heuman's Lambda function, $\Lambda_0(\psi,p)$, (see Appendix A).

Using (\ref{piapprox}) it is straightforward to find the following approximate expressions

\bea
\label{tetaapprox}\Delta\theta&\approx&\frac{\pi}{2}\ \{\frac{\sqrt{u_c+1}}{\sqrt{u_c-1}}-1\}\\
\label{Japprox}J&\approx&n\ \frac{\sqrt{\lambda}}{4}\  u_l\\
\label{EJapprox}E-\omega J&\approx& \frac{2n}{2\pi}\ \sqrt{\lambda}\ \frac{2E(1/\sqrt{2})-K(1/\sqrt{2})}{\sqrt{u_c-1}}\ {u_l}^{1/2}
\eea

Recalling that the number of spikes is given by $n=\pi/{\Delta\theta}$, it is evident that the limit we are
considering describes strings with large angular momentum and fixed number of spikes. This number decreases as
$\rho_c$ approaches zero $(u_c\rightarrow1)$. As $\rho_c$ gets smaller, the angle covered by a single string segment increases
and in the limiting case it blows up resulting in a spiral with infinite energy, like the one
we encountered in the flat space, and is not a physically allowed configuration (See the footnote (\ref{foot})).

\begin{figure}[ht]
\begin{center}
\includegraphics[scale=0.7]{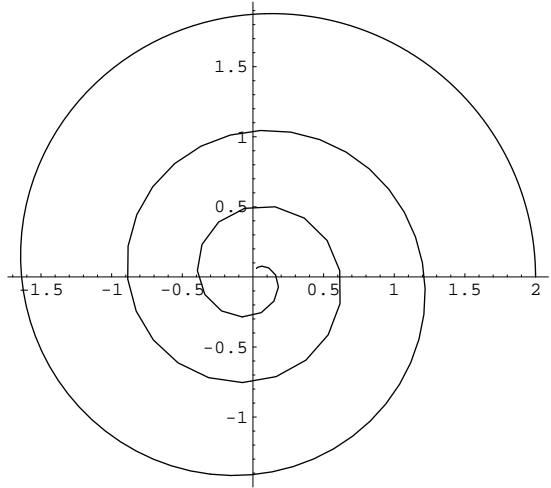}
\caption{A string segment forming a spiral. $\rho_c$ is so small that covers an angle larger than the minimal allowed value, $\pi/2$.}
\end{center}
\end{figure}

We may thus assume that $\rho_c$ is much larger than one, to avoid the above mentioned situation.
In this limit $\omega\approx1$ and we find the following
expressions, which this time we write in terms of $\rho$ instead of $u$

\bea
\Delta\theta&\approx&\pi\ e^{-2\rho_c}\\
J&\approx&n\ \frac{\sqrt{\lambda}}{8}\  e^{2\rho_l}\\
\label{EJapprox2}E-J&\approx& n\ \frac{\sqrt{\lambda}}{\pi}\ [2E(1/\sqrt{2})-K(1/\sqrt{2})]\ e^{(\rho_l-\rho_c)}
\eea

Looking at (\ref{EJapprox2}), we require $\rho_l\gg\rho_c$ such that $(E-J)/n$ remains large
and a semi classical approximation for each spike remains valid. This means that although we
require $\rho_c$ to be large enough to produce a finite number of spikes, we don't want
it to be so large to make each spike infinitesimal in size and energy.

Gathering all the relations above we can write the dispersion relation for spiky strings in
the limit of large angular momentum

\bea
\label{EJspike}
E&\approx& J+4n\ \sqrt{\lambda}\ \frac{2E(1/\sqrt{2})-K(1/\sqrt{2})}{\sqrt{2}\pi}\
\left(\frac{1}{n}\frac{J}{\sqrt{\lambda}}\right)^{1/2}\nonumber\\
&\approx&J+1.34\ \sqrt{\lambda}\ \frac{n}{2\pi}\ \left(\frac{4\pi}{n}\frac{J}{\sqrt{\lambda}}\right)^{1/2}
\eea

where in the last line we have plugged the approximate values of the elliptic integrals.
This is to be compared with the relation obtained for the spiky strings found in \cite{Kkk}

\be
E\approx J+\sqrt{\lambda}\ \frac{n}{2\pi}\ \ln\left(\frac{4\pi}{n}\frac{J}{\sqrt{\lambda}}\right)
\ee

As can be seen, apart from a numerical factor of order of unity,
the dispersion relation for dual spiky strings differs from
the one for the usual spikes. This is because the dual spikes near the boundary
look like portions of almost circular strings rather than folded ones. This changes
the anomalous dimension from the usual logarithm dependence to square root
which is the known behavior for rotating and pulsating circular strings in AdS.
In the discussion section we will comment more on this point.


\section{Dual spikes on sphere}

In this section we find spiky string solutions on $Ads_5\times S^5$ which live
in the origin of $AdS$ and which are point-like in $S^3\subset S^5$ i.e. they live
on a hemisphere with the following metric

\be
dS^2=-dt^2+\cos^2\theta d\phi^2+d\theta^2
\ee

where $0\le\theta\le\pi/2$ and $0\le\phi\le\pi$.

In \cite{Ryang}, such solutions with spikes pointing towards the equator, $\theta=0$, were found.
In the following we find solutions with spikes pointing towards the pole $\theta=\pi/2$. The ansatz we consider is

\be
t=\tau\ ,\ \ \ \ \ \phi=\omega t+\sigma\ ,\ \ \ \ \ \theta=\theta(\sigma)
\ee

The Nambu-Gotto Lagrangian for this ansatz will read

\[
{\cal L}_{NG}=-\frac{\lambda}{2\pi}\sqrt{l}\nonumber
\]

with

\be
l=(1-\omega^2\cos^2\theta)\ \theta'^2+\cos^2\theta
\ee

the isometry direction $\partial_{\phi}$ results in a constant of motion,
which we denote by $\theta_l$ and is given by

\be
\label{thetalobe}
\cos^2\theta_l\equiv\ \frac{\cos^2\theta}{\sqrt{l}}
\ee

As we will see below, this constant determines the position of lobes in the string configuration.
The other free parameter of the problem, $\omega$, determines the position of cusps by $\cos\theta_c=1/\omega$.
From (\ref{thetalobe}) we find

\be
\frac{\theta'}{\cos\theta}=\frac{\cos\theta_c}{\cos\theta_l}\ \sqrt{\frac{\cos^2\theta_l-\cos^2\theta}
{\cos^2\theta-\cos^2\theta_c}}
\ee

Here again one can check that if the above relation holds the equations of motion
are also satisfied. In \cite{Ryang} it was assumed that $\theta_c<\theta_l$. Here we make the assumption
$\theta_l<\theta_c$ which will lead to a second class of spiky solutions which we call
dual spiky strings. These strings are formed by attaching $2n$ number of string segments
which stretch between $\theta_l$ and $\theta_c$. The angle covered by each segment, $\Delta\phi$, and
also the energy, $E$, and angular momentum, $J$, of the closed string are given by the following integrals

\bea
\Delta\phi&=& \frac{\cos\theta_l}{\cos\theta_c}\int_{\theta_l}^{\theta_c}\frac{d\theta}{\cos\theta}
\frac{\sqrt{\cos^2\theta-\cos^2\theta_c}}{\sqrt{\cos^2\theta_l-\cos^2\theta}}\\
J&=&\sqrt{\lambda}\frac{2n}{2\pi}\int_{\theta_l}^{\theta_c}d\theta\cos\theta
\frac{\sqrt{\cos^2\theta_l-\cos^2\theta}}{\sqrt{\cos^2\theta-\cos^2\theta_c}}\\
E-\omega J&=&\sqrt{\lambda}\frac{2n}{2\pi}\omega \int_{\theta_l}^{\theta_c}d\theta\cos\theta
\frac{\sqrt{\cos^2\theta-\cos^2\theta_c}}{\sqrt{\cos^2\theta_l-\cos^2\theta}}
\eea

The integrals can be computed with the following results

\bea
\Delta\phi&=&\frac{x_c\ q^2}{\sqrt{(1-x_c^2)(1-x_l^2)}}\ K(q)-\frac{\pi}{2}[1-\Lambda_0(\psi,q)]\\
J&=&\sqrt{\lambda}\ \frac{2n}{2\pi}x_c\ [E(q)-\frac{x_l^2}{x_c^2}\ K(q)]\\
E-\omega J&=&\sqrt{\lambda}\ \frac{2n}{2\pi}\frac{x_c\ q^2}{\sqrt{1-x_c^2}}\ [K(q)-E(q)]
\eea

where

\[
x=\sin\theta\ ,\ \ \ \ \ q^2=1-\frac{x_l^2}{x_c^2}\ ,\ \ \ \ \ \psi=\sin^{-1}\left(\frac{1-x_c^2}{1-x_l^2}\right)^{1/2}
\]

One can see that for this class of spiky strings the angular momentum can never be large.
Therefore there is no limit where the world sheet corrections to the classical
configuration become sub dominant and a semi classical analysis becomes valid.
Despite this, and in order to simplify the expressions,
we consider some limits of the parameters in what follows.

First consider the case when $x_l\ll 1$ and $x_c$ is fixed. In this limit $q\approx1$ and

\[
K(q)\approx\ln(\frac{1}{\sqrt{1-q^2}})\ ,\ \ \ \ \ E(q)\approx1\ ,\ \ \ \ \ \frac{\pi}{2}\Lambda_0(\psi,q)\approx\psi
\approx\frac{\pi}{2}-\theta_c
\]

Using the above approximate expressions we can find that
the limit $q\approx1$ corresponds to fixed angular momentum and large energy for each half spike.
A string segment stretching between $\theta_l\approx0$ and $\theta_c$ looks like a spiral which covers
a large angle. The limit therefore does not correspond to a physically valid configuration
\footnote
{\label{foot}This configuration
is not physical in the sense that it is not compatible with our original ansatz for the string where we
assumed a winding number $w=1$ for the string. Including a general winding number in the ansatz
might give rise to interesting new configurations including ones made up of spirals. In fact the
``single spikes" of \cite{Ishizeki:2007we} which live on sphere are of this kind. For these solutions
the infinite winding number is replacing $J$ in the dispersion relation to give a finite
anomalous part. The interchange of angular momentum and winding number is reminiscent of T-duality
relations. The dual spikes were shown to be T-duals of the usual ones in flat space.
Such a relation in AdS and sphere between spikes and dual spikes might be an interesting
subject to investigate.}.


Next consider a nearly circular configuration with  $x_c\approx x_l$ or $q\approx0$.
For this case we have

\[
K(q)\approx\frac{\pi}{2}(1+\frac{\epsilon}{2x_c})\ ,\ \ \ E(q)\approx\frac{\pi}{2}(1-\frac{\epsilon}{2x_c})\ , \ \ \
\Lambda_0(\psi,q)\approx1-\sqrt{\frac{2}{1-x_c^2}}\ \epsilon\ ,\ \ \ \ \ \ (\epsilon\equiv x_c-x_l)
\]

We therefore find

\bea
\Delta\phi&\approx&\sqrt{\frac{2}{1-x_c^2}}\left(\sqrt{\frac{2}{1-x_c^2}}-1\right)\frac{\pi}{2}\epsilon\\
J&\approx&\sqrt{\lambda}\ \frac{n}{2}\ \epsilon\\
E-\omega J&\approx& \sqrt{\lambda}\ \frac{n}{2}\ \frac{\epsilon}{\sqrt{1-x_c^2}}
\eea

This limit thus describes a large number of half spikes, each with a small energy and angular momentum,
resulting in a closed string with a finite $E$ and $J$.

If we further assume that the string is close to the equator, $x_c\approx x_l\approx0$ and hence
$\omega\approx1$, we find

\be
E\approx 2J
\ee

This coincides with the expression found in \cite{Ryang}
for nearly circular spiky strings near the equator but of course, and unlike our case,
with spikes pointing towards the equator.

Now consider the string to be near the pole, $x_c\approx x_l\approx1$ and hence $\omega\rightarrow\infty$.
To find the dispersion relation we should first express $x_c$ in terms of the charges which, to
the leading order, turns out to be

\[
1-x_c^2\approx n\epsilon\approx \frac{2J}{\sqrt{\lambda}}
\]

As a result we find

\be
E^2\approx 2\sqrt{\lambda}\ J
\ee

This differs from the result $E^2\approx4\sqrt{\lambda}(\sqrt{\lambda}+14/3\ J/n)$ found in \cite{Ryang} for
nearly circular spiky strings close to the pole and with spikes pointing towards the equator.


\section{Discussion}

Here we are mainly interested in the field theory interpretation of dual spikes
in AdS, namely the result (\ref{EJspike}) (remember that we did not have a large $J$ limit
for dual spikes on sphere which makes the semi classical limit difficult to interpret in
field theory). As was mentioned earlier, the result (\ref{EJspike}) looks like that of
circular rotating and pulsating strings \cite{park}. This is in fact understandable
as we will discuss below.

Semi classical string configurations in AdS have led to the picture that spikes on
string represent fields in SYM whereas a profile in the radius of AdS, which
should generically be accompanied by rotation to give a string solution, are
represented by covariant derivatives $D_+$ which also carry spin.
The logarithm behavior of anomalous dimension for spinning strings is believed
to be caused by the large number of derivatives as compared to fields.

Circular rotating and pulsating strings\footnote{
For circular rotating and/or pulsating strings see for example\cite{Minahan:2002rc},\cite{park}.},
on the other hand, have been proposed to be represented
by self dual gauge field strengths $F^{(+)}$ (see \cite{park}). These configurations have equal rotations
$S_1=S_2$ in the orthogonal planes of $S^3$ to stabilize and thus carry $(1,0)$ representation of
$SO(4)$ as $F^{(+)}$. The anomalous dimension for mostly spinning strings, in these solutions,
is like $(\lambda S)^{1/3}$ and is like $\lambda^{1/4}(N)^{1/2}$ for mostly pulsating ones
where $N$ is the oscillation number for pulsation.

The dual spikes we have studies in this paper, in the large $J$ limit and near
the boundary, look like portions of circular rotating strings. Moreover,
the profile in $\rho$ in addition to rotation induce a pulsating motion
for the string as seen by an observer in fixed angle. In fact
in the large $J$ limit of the dual spikes, letting $\eta\equiv\omega-1$ and
keeping $\eta$ small (with $\eta J$ fixed) to keep the number of spikes finite, we get

\[
E\approx J+\eta J+1.34\ \sqrt{\frac{2}{\pi}}\ n\ \lambda^{1/4}\ \sqrt{\eta J}
\]

Here $\eta J$ is replacing the oscillator number for pulsation. For smaller $\eta$
the portion of string near the boundary becomes more circular which reduces the
pulsation-like movement of the string and gives a smaller oscillation number.
One might then guess that the dual spiky strings in AdS in the large angular
momentum limit are schematically represented by operators of the form

\[
{\cal O}\sim {\bf Tr}\{\Pi_i^n (D_+)^{J/n} (D_+ D_-)^{\eta J/2n} \ \Phi_i\}
\]

The $D_+$ operators are responsible for the profile in $\rho$ as well as the rotation. The $D_+D_-$, on the
other hand, contribute to the dimension but not to the angular momentum. The fields $\Phi$ as before
represent the spikes on the string.
It is clear that a better understanding of the field theory interpretation for
these solutions needs a more careful study which we postpone to a future work.

It might also be very interesting, though seemingly difficult,
to find spiky solutions with a real pulsation, perhaps interpolating
between spikes and dual spikes.

\section*{Acknowledgements}
AEM would like to thank H. Arfaei for discussions and M.M. Sheikh-Jabbari
for discussions and comments on the draft and also M.R. Maktabdaran
for helping with the figures. We would especially like to thank M. Kruczenski for useful
discussions and comments on the draft.

\section*{Appendix A: Some useful relations for Elliptic integrals}\label{appendix}

Here we gather some relations regarding Elliptic integrals. For a complete account see \cite{grad},\cite{abram}.

The Elliptical integrals of first, second and third kind, $F$, $E$ and $\Pi$ are defined as
\bea
F(\alpha;q)&=&\int_0^\alpha d\theta \frac{1}{(1-q \sin^2\theta)^{\frac{1}{2}}}\\
E(\alpha;q)&=&\int_0^\alpha d\theta (1-q \sin^2\theta)^{\frac{1}{2}}\\
\Pi(\alpha;n,q)&=& \int_0^\alpha d\theta {\frac{1}{(1-n
\sin^2\theta)(1-q^2 \sin^2\theta)^{\frac{1}{2}}}}
\eea

Complete Elliptic integrals are those with $\alpha=\pi/2$. A usual notation is

\[
K(q)\equiv F({\frac{\pi}{2}},q)
\]

For the limiting case $q\rightarrow 0$ we have

\[
K(q)\approx{\frac{\pi}{2}}\ (1+\frac{q^2}{4}),\ \ \ \ \ E(0)\approx{\frac{\pi}{2}} (1-\frac{q^2}{4})
\]

In the other
limiting case $q\rightarrow 1$
\[
K(q)\approx \ln\frac{4}{q'}\ ,\ \ \ \ \ E(q)\approx 1+\frac{q'^2}{2}\ln\frac{4}{q'}\ ,\ \ \ \ \ (q'\equiv\sqrt{1-q^2})
\]

For the complete Elliptic integrals of the third kind, $\Pi(n,q)$, the cases $0<n<q^2$ and $n>1$ are known
as ``hyperbolic" whereas those with $q^2<n<1$ and $n<0$ are known as``circular" cases. For $q^2< n <1$ we have
the following relation

\[
\Pi(n,q)=K(q)+\frac{\pi}{2}\delta[1-\Lambda_0(\psi,q)]
\]

where

\[
\delta=[{\frac{n}{(1-n)(n-q^2)}}]^{\frac{1}{2}} ,\ \ \ \ \
\psi=\arcsin[{\frac{1-n}{1-q^2}}]^{\frac{1}{2}}
\]

and $\Lambda_0$ is Heuman's lambda function defined as

\[
\Lambda_0(\psi,q)={\frac{2}
{\pi}}\left(K(q)E(\psi;q')-[K(q)-E(q)]F(\psi;q')\right)
\]

For $n<0$ we have

\[
\Pi(n,q)={\frac{-n(1-q^2)}{(1-n)(q^2-n)}}\pi(N,q)+{\frac{q^2}{(q^2-n)}}K(q)
\]

where $N={\frac{q^2-n}{1-n}}$.

\end{document}